\documentclass[rsi,amsmath,amssymb,reprint]{revtex4-1}
\usepackage{graphicx}
\usepackage{dcolumn}
\usepackage{bm}

\usepackage[utf8]{inputenc}
\usepackage[T1]{fontenc}
\usepackage{mathptmx}
\usepackage{etoolbox}
\usepackage{ctable}

\makeatletter
\def\@email#1#2{%
 \endgroup
 \patchcmd{\titleblock@produce}
  {\frontmatter@RRAPformat}
  {\frontmatter@RRAPformat{\produce@RRAP{*#1\href{mailto:#2}{#2}}}\frontmatter@RRAPformat}
  {}{}
}%
\makeatother
\begin{document}

\title{X-ray Microscopy and Talbot  Imaging with the Matter in Extreme Conditions X-ray Imager at LCLS}

\author{
Eric Galtier\textsuperscript{1},
Dimitri Khaghani\textsuperscript{1},
Nina Boiadjieva\textsuperscript{1},
Mikako Makita\textsuperscript{2},
Arianna E. Gleason\textsuperscript{1},
Silvia Pandolfi \textsuperscript{1,3},
Anne Sakdinawat \textsuperscript{1},
Yanwei Liu \textsuperscript{1},
Daniel Hodge \textsuperscript{4},
Richard Sandberg\textsuperscript{4},
Gilliss Dyer\textsuperscript{1},
Phil Heimann\textsuperscript{1},
Frank Seiboth\textsuperscript{5},
Hae Ja Lee\textsuperscript{1},
Bob Nagler\textsuperscript{1}
}

\affiliation{
1. SLAC National Accelerator Laboratory, 2575 Sand Hill Rd., Menlo Park, CA 94025, USA.\\
2. European XFEL, Holzkoppel 4, 22869 Schenefeld, Germany
3. Sorbonne University, Paris, France
4.Brigham Young University, Department of Physics and Astronomy, Provo, UT, 84602, USA
5. Center for X-ray and Nano Science, Deutsches Elektronen-Synchrotron DESY, Notkestraße 85, DE-22607 Hamburg, Germany
}

\date{\today}
\begin{abstract}
The last decade has shown the great potential that  X-ray Free Electron Lasers (FEL) have to study High Energy Density matter. Experiments at FELs have made significant breakthroughs in Shock Physics and Dynamic Diffraction,  Dense Plasma Physics and Warm Dense Matter Science, using techniques such as isochoric heating, inelastic scattering, small angle scattering and x-ray diffraction. In addition, and complementary to these techniques, the coherent properties of the FEL beam can be used to image HED samples with high fidelity. We present  new imaging diagnostics and techniques developed at the Matter in Extreme Conditions (MEC) instrument at Linac Coherent Light Source (LCLS) over the last few years. We  show result of a previously used Phase Contrast Imaging setup, where the  X-ray beam propagates from the target to a camera some distance away revealing its phase, as well as  a  imaging approach where the target is re-imaged on the camera with 300nm resolution. Last, we show a new Talbot Imaging method allowing  both  x-ray phase and intensity measurements change introduced by a target with sub-micron resolution.

\end{abstract}

\maketitle

\section{Introduction}
High Energy Density (HED) science has traditionally made ample use of X-ray radiography and imaging to  investigate a variety of phenomena, such as the implosions of inertial confinement fusion (ICF) capsules~\cite{Katayama1993,Kalantar1997,Marshall2009,Hicks2010,Dewald2018,Weber2022}, X-pinch plasmas~\cite{Shelkovenko2001} and the general hydrodynamics of evolution of HED targets.~\cite{Landen2001,Oliver2022}.
Since the dynamics of these phenomena evolve on very fast time scales (from fs to ns), ultrashort X-ray pulses are needed to obtain relevant data. Historically,  X-ray sources generated by the interaction of high power laser with backlighter targets have been employed, with typical spatial and temporal resolution on the order of tens of micrometers and hundreds of picoseconds or more~\cite{Landen2001,Morace2014}
Phase Contrast Imaging techniques with backlighters have also been developed~\cite{Montgomery2004,Kozioziemski2005,Koch2009,Ping2011}, albeit with limited resolutions compared to synchrotrons, where sub-micron resolutions are standard~\cite{Nugent1996,Cloetens1999}.
 
The start of operation of LCLS~\cite{Emma2010} in 2009  opened new possibilities for X-ray imaging in HED science. The coherent properties of the beam allow for imaging with sub-micron spatial resolution,  while the short, bright pulses allow temporal resolution of tens of femtoseconds. These combined properties allow for single-shot X-ray imaging techniques to emerge as a leading technique to further our understanding of complex HED phenomena. At LCLS, the Matter in Extreme Conditions end-station~\cite{Nagler2015} is specifically tailored to field experiments in HED science. To foster this field of research, the different imaging techniques and instruments have been developed, and are available to its user community.

In this paper, we describe the use and performances of the MEC Xray Imager (MXI) and the different diagnostic imaging techniques that make use of it, and present  preliminary proof of principle experiments to showcase its potential for HED science.  

\section{Instrument overview}
\label{sec:overview}

\begin{figure}
\includegraphics[width=\linewidth]{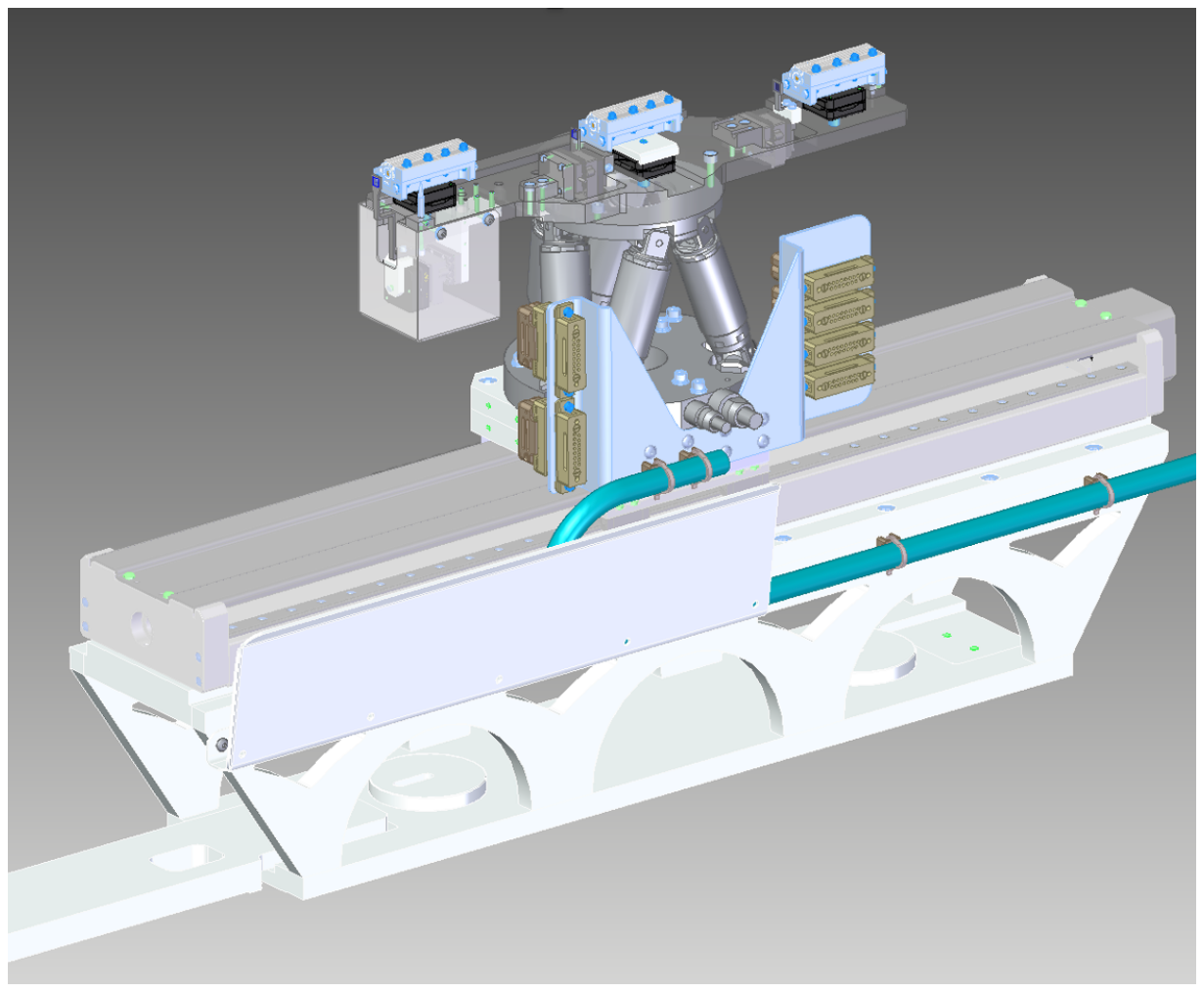}
\caption{\label{fig:mxi_cad} 3D model of the MXI assembly.} 
\end{figure}

The MXI instrument is designed for use in the MEC vacuum chamber, details of which can be found in~\cite{Nagler2015}. 
It uses compound refractive Beryllium lenses to focus the X-ray beam to sub-micron spot sizes. The capabilities of such lens stacks have been described extensively in the literature~\cite{Snigirev1996,Lengeler1999a,Lengeler1999b,Lengeler2002}, and have been used for this purpose at MEC~\cite{nagler2016,Schropp2012,seiboth2017}. A stack of lenses is placed on alignment stages that positions the lenses in the beam. An 3D model is shown in Figure~\ref{fig:mxi_cad}.  Up to three stacks of lenses can be placed on the instrument simultaneously, to allow usage of different photon energies and/or focal lengths during an experiment without venting the MEC vacuum chamber. The front stack can accommodate up to 100 lenses, necessary for the higher photon energies, while the other two stacks are limited to 40 lenses. The lens stacks are mounted on a hexapod, allowing for precise alignment with six degrees of freedom, and a translation stage of 500\,mm to change the distance between the lens stack and the object plane. Since stacks of Be lenses are known to have aberrations, which can limit their use in imaging and phase contrast application, a phase plate that corrects these aberrations\cite{seiboth2017} can be placed 10\,mm in front of the lens set. The phase plate in aligned in the horizontal and vertical direction with two small linear stages, which can also completely retract the corrector. The MXI lens sets can be placed either upstream of the sample (see Sec.\ref{sec:pci}) for phase contrast imaging or downstream (see Sec.\ref{sec:di}) for direct imaging of a sample.

\begin{figure}
\includegraphics[width=\linewidth]{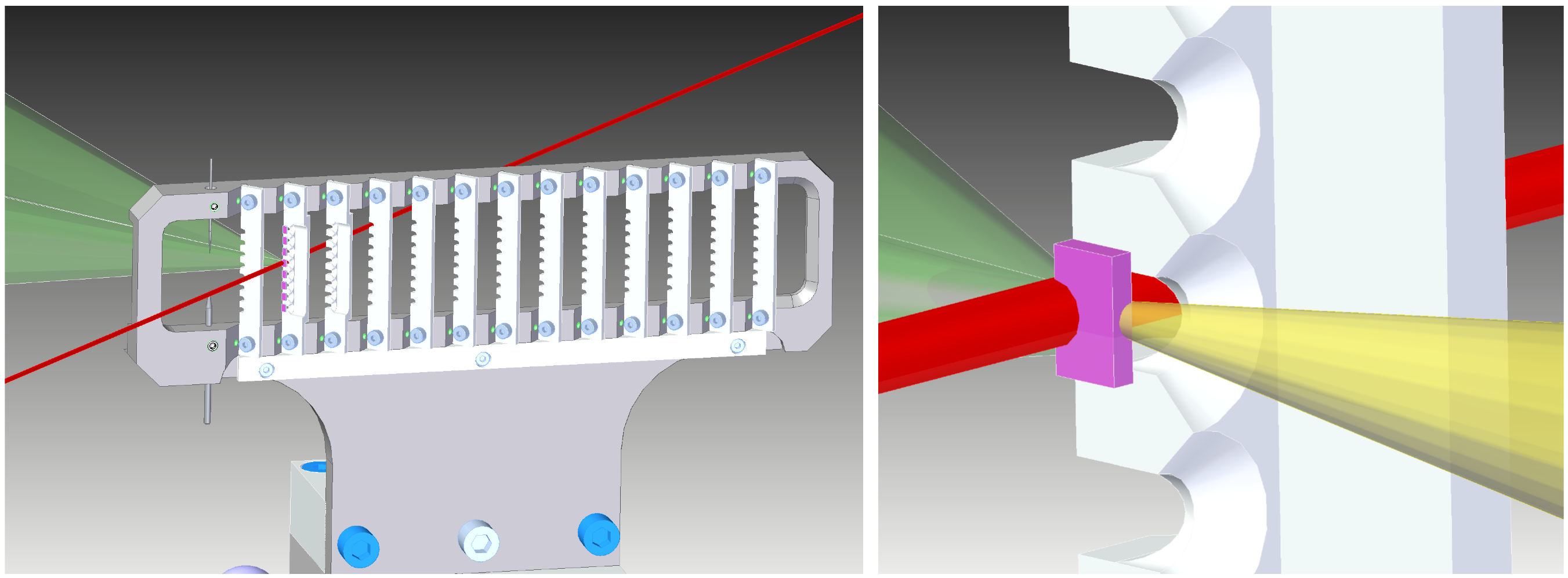}
\caption{\label{fig:targetholder} left) 3D model of the target holder with an orthogonal configuration between the long pulse drive laser (in green) and the X-rays (in red). right) zoom in of a single target an a pillar suitable for use with the MEC VISAR diagnostic~\cite{Nagler2015} (VISAR beam in yellow).}
\end{figure}

In a series of experiments, samples are driven by the MEC short or long pulse laser orthogonally to the imaging direction (e.g., a shock is imaged traveling through a sample). To this end a sample holder allowing such geometry is available (see Fig.~\ref{fig:targetholder}). 
The samples are aligned using two long distance microscopes (see~\cite{Nagler2015} for details), which view the sample at orthogonal directions, and allow for precise and repeatable alignment. The samples are mounted on a pillar that is screwed to the sample holder. For use in shock experiment, specialty pillars that allow for VISAR~\cite{Barker79} measurements on thin samples are available.

The image is recorded on an X-ray camera that is placed behind the MEC chamber. The camera can be placed between 1.3m to 4.3m behind the sample plane. In general, the farthest distance is preferred to have the largest magnification, although some application may benefit from the shorter distance (i.e., ptychography~\cite{Schropp2010}). For most experiments, we have used an Optique Peter X-ray microscope, consisting of a Ce:YAG scintillator and optical microscope, coupled to a Andor Zyla optical camera, which can have a resolution of ~2 $\mu$m. Direct imaging cameras (e.g., Epix or Jungfrau) can be used instead, however the resolution is limited  by the pixel size (typically 50$\mu$m for Epix10k and 75um for Jungfrau). In addition, an Icarus V2, an ultra-fast x-ray imager (UXI) camera with pixel size of 25$\mu$m,~\cite{Hart2019,Looker2020,Hodge2022} and  capable of capturing  multiple image frames  spaced as little as 2ns apart, can also be used, in conjuction with the LCLS multibunch operation mode~\cite{Decker}. An advantage of direct detection compared to scintillator based detection is that the camera's are orders of magnitude more sensitive, and therefore can be used with thicker, higher Z, more absorbent samples. 


\section{Phase Contrast Imaging geometry}
\label{sec:pci}
\begin{figure}
\includegraphics[width=\linewidth]{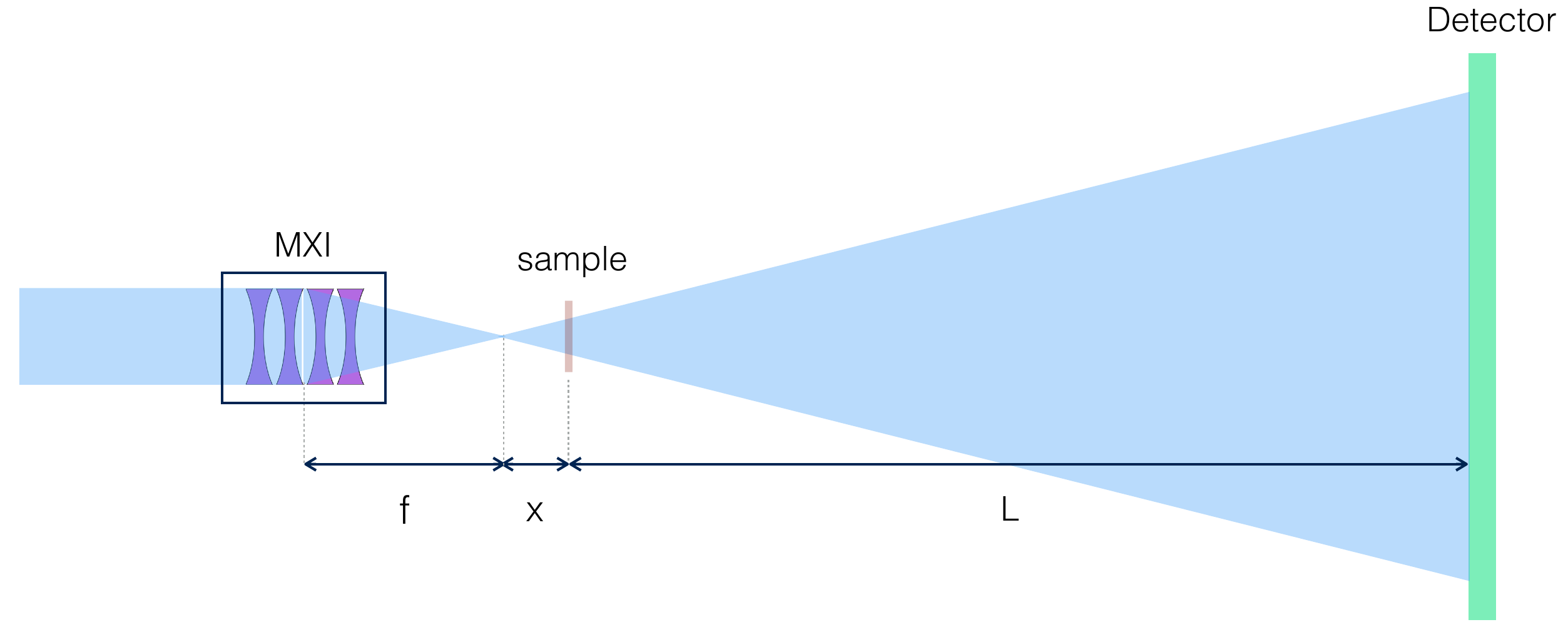}
\caption{\label{fig:pcigeo} Sketch of the MXI setup in phase contrast imaging geometry.} 
\end{figure}

\begin{figure}
\includegraphics[width=\linewidth]{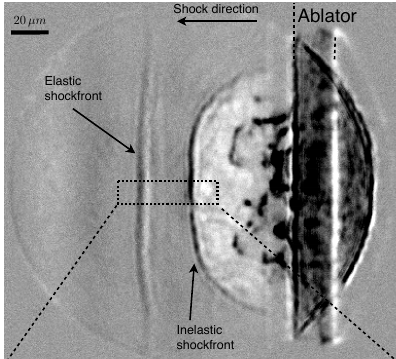}
\caption{\label{fig:pcishock}  The MEC long pulse laser impinges on the a plastic ablator coated on a Si target. An elastic shockfont, followed by a second shockfront of a phase transition travels from right to left.} 
\end{figure}

The imaging geometry first deployed at MEC~\cite{Schropp2015} used a diverging beam at target and optic-free propagation from target to detector plane. A schematic of the setup can be seen in Fig.\ref{fig:pcigeo}.
The MXI, with a lens set of focal length $f$, is placed in the MEC vacuum chamberd and focuses the LCLS beam  to approximately 100nm. A certain distance, x, behind the focus, a sample is placed in the divergent beam. The x-ray beam propagates to  a detector at a distance L behind the sample, leading to a geometric magnification of the sample of L/x. Due to the long propagation distance of the X-ray, phase difference induced by the sample can readily be seen in the image (e.g., clear fringes appear at sharp phase boundaries). For this reason this geometry is called the Phase Constrast Imaging (PCI) geometry at MEC. Iterative algorithms can be used to retrieve the phase, and if this is successful, the full complex transmission function (i.e., absorption and phase shift) can be calculated. Experiment in this geometries have been fielded at MEC, yielding scientific results on shockwave propagation in diamond~\cite{Schropp2015} and the kinetics of phase transformations in silicon~\cite{Brown2019}.
The setup can be easily combined with X-ray diffraction. Moving the lenses of the MXI closer to the target reduced the field of view of the image, and therefore the area where X-rays hit the target. In this way, diffraction from an area of only a few microns in size can be obtained, and different crystallographic phases in the sample can be identified~\cite{Brown2019}. A typical image of a laser-generated shock wave travelling through a sample can be seen in Fig.~\ref{fig:pcishock} and  more examples have been published in~e.g.,~\cite{Schropp2015,Nagler2015,nagler2016}

\section{Direct Imaging geometry}
\label{sec:di}
\begin{figure}
\includegraphics[width=\linewidth]{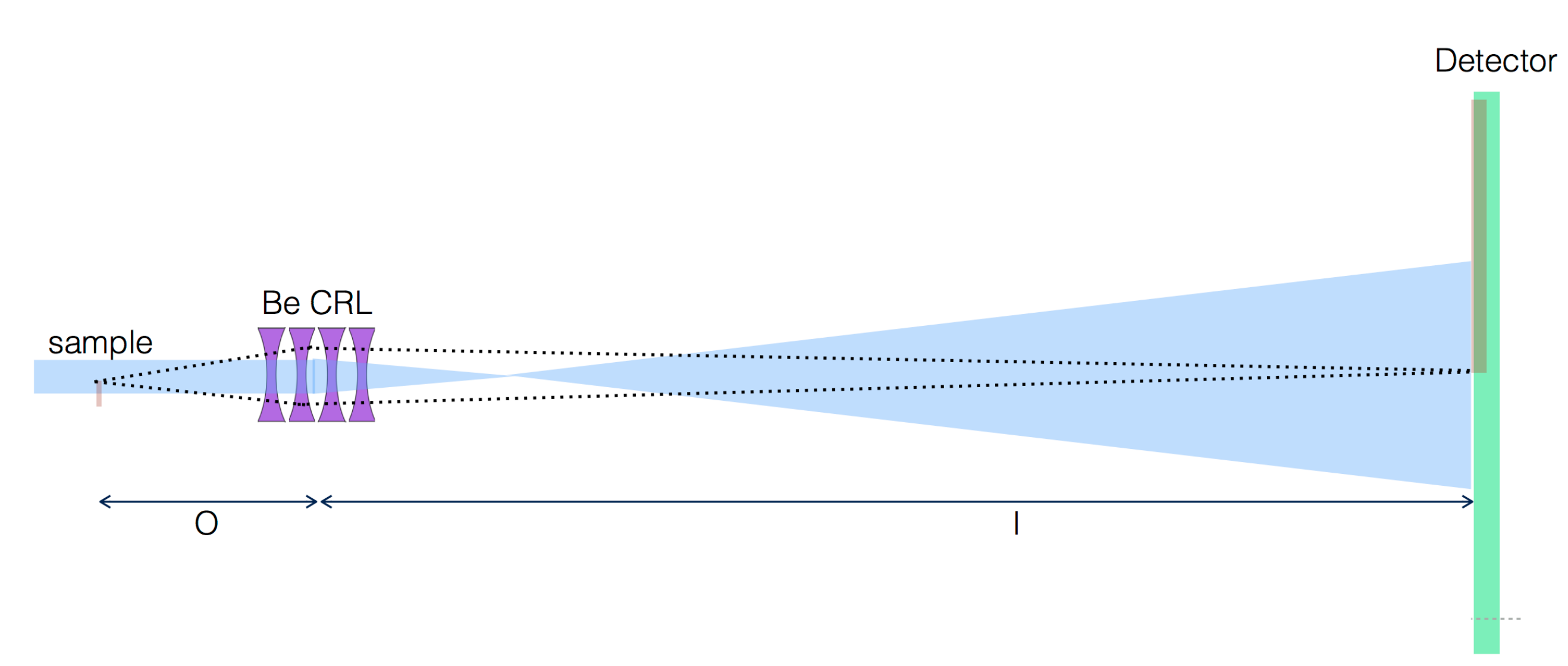}
\caption{\label{fig:directimaging} MXI in direct imaging geometry} 
\end{figure}

While the PCI geometry described above has great potential when a full phase retrieval can be performed, this has been in most cases elusive at MEC.  Without such a phase retrieval, the resolution can be severely limited, and an alternative setup that directly images a sample on the detector has been developed and used at MEC. A sketch of the setup is shown in Figure~\ref{fig:directimaging}.

\begin{figure}
\includegraphics[width=\linewidth]{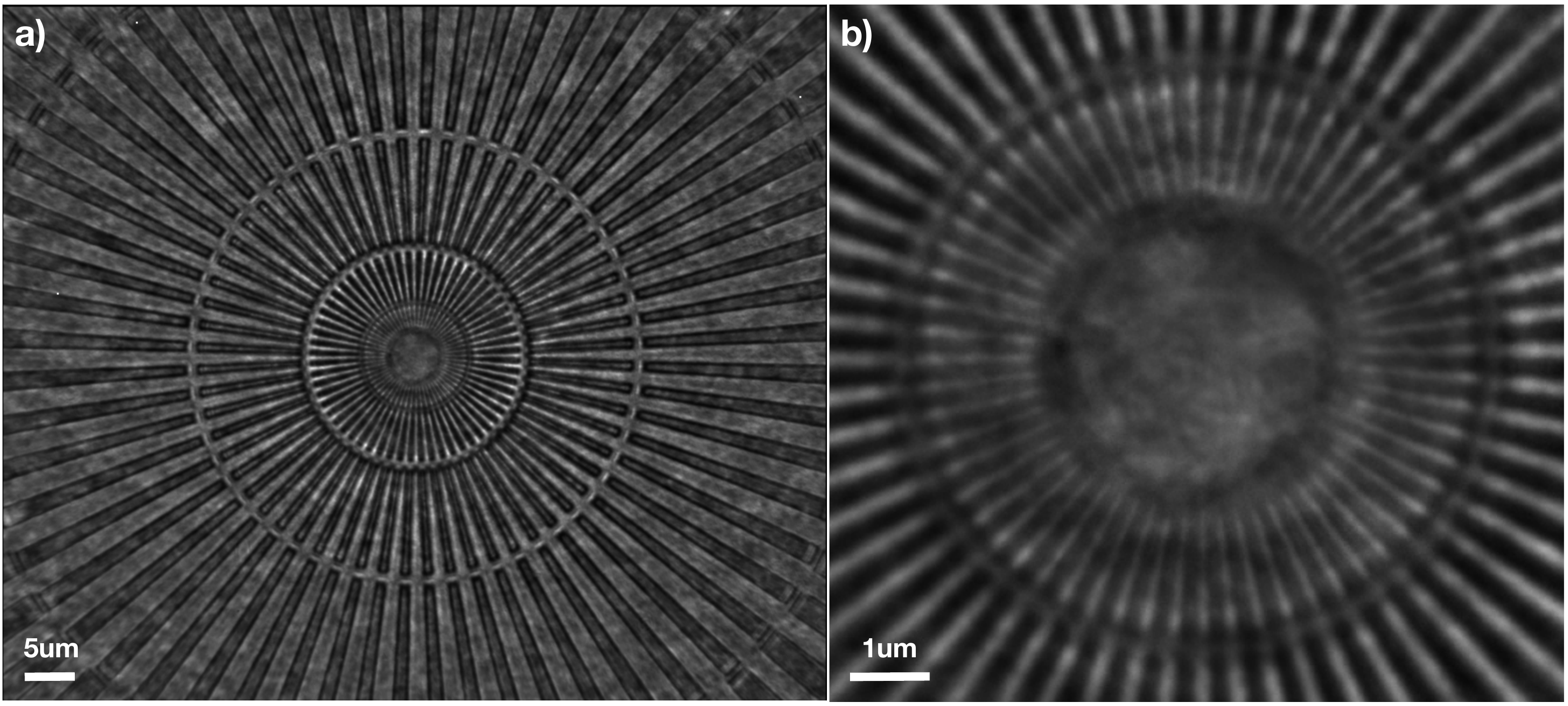}
\caption{\label{fig:res8keV} 
Imaging of a Siemens star resolution target at 9.5keV in a false color scale. a)Full field of view.  b) zoom of the center. Smallest features have a spacing of 200nm. } 
\end{figure}

In this geometry the Be lenses are placed behind the sample such that the distance to the sample, $O$,  and the distance to the camera, $I$,  matches the well know lens formula $O^{-1}+I^{-1}=f^{-1}$, with $f$ the focal length of the Be lenses. Such a setup allows for imaging intensity contrast of the sample with sub-micron resolution, even when phase reconstruction is not possible. 
Result showing a resolution better than 250\,nm  are shown in Figure~\ref{fig:res8keV}. 

\begin{figure}
\includegraphics[width=\linewidth]{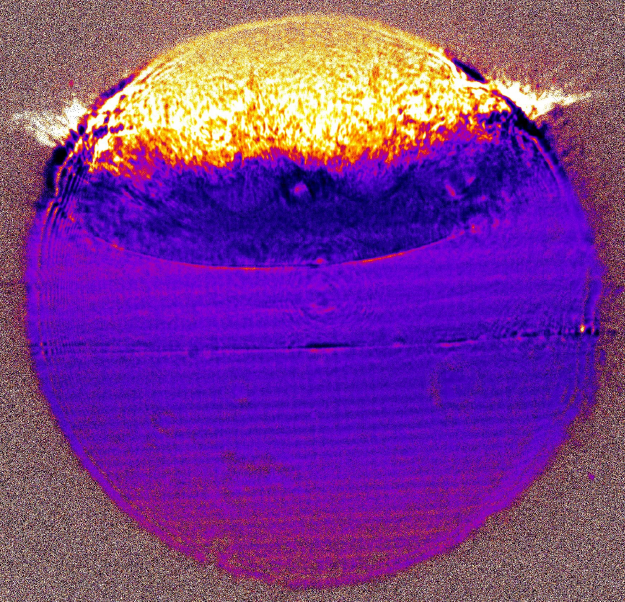}
\caption{\label{fig:shock_Si} 
White-Field corrected image of a shock wave propagating through a silicon sample. The MEC long pulse laser, with pulse length 10\,ns and  energy 50\,J was focused with continuous phase plates to a top-hat profile with 150$\mu$m diameter on the side of a Si target. Image is taken 10\,ns after the laser strikes the target. Both the elastic precursor shock, and the phase transition wave can be clearly distinguished.} 
\end{figure}

However, imaging samples in this way with beryllium lenses is far from ideal.
First and foremost the image of the x-rays on the detector when no sample is present has significant structure that are introduced by the Be lens stack. Some of this can be treated by a simple white field correction (i.e., dividing the image of the sample by the white field of image when no sample is present) or by more sophisticated analysis method such as Principal Component Analysis~\cite{Hagemann2021,Hodge2022}. This is the analysis that is performed in the images in Figures~\ref{fig:pcishock}, \ref{fig:res8keV} and \ref{fig:shock_Si}. However, this method does 
not work very well in areas of the image where strong phase changes are expected (e.g. at a shock front). This effect gets worse when lenses upstream of the MEC target chamber are placed into the beam to pre-focus the x-rays onto the part of the sample that is imaged, to increase signal level. 
Furthermore, both wavefront aberrations and chromatic aberrations are present in the imaging lens stack.  The chromatic aberration can be mitigated by a monochromator, however this is not always feasible: the loss of photons due to the monochromator can reduce the number of photons below what is necessary to get an adequate  signal on the camera. The wavefront aberrations can be reduced with the appropriate phase-correcter if it is available; however this correction is never perfect. Furthermore, the limited aperture of the lenses give an intrinsic limit to the imaging that can be achieved. Since the LCLS light source is virtually fully spatially coherent, these imaging imperfection tend to show up as fringes, especially around abrupt changes in density where jumps in the phase of the x-rays are created (e.g., sample edges, edges of wires, shock front, etc.).  Furthermore, samples can be thicker than the theoretical Rayleigh length of the focus (or depth of focus of the imaging), leading to significant propagation effects inside the sample. The commonly used assumption that the sample can be modelled by an infinitely thin absorption and phase masks breaks down. These effects can  present significant challenges to the interpretation and modelling of the data. A full treatment of these analysis challenges is outside the scope of this paper.

The MXI instrument was tested in the direct imaging geometry at 8.2\,keV  and 18\,keV photon energy. For 8.2\,keV, a set of 25 Beryllium lenses with a curvature of 50\,$\mu$m were used. The set has a focal length of 204\,mm and was placed  214\,mm downstream of the target, leading to an image at the detector, placed  4.1\,m behind the lenses, with a magnification of  approximately 20. The position of the lenses was tuned to get the best imaging resolution, using resolution targets. Images of the resolution target are shown in Fig.\ref{fig:res8keV}. Line widths  as small as 250\,nm can be clearly resolved.

At photon energies as high as 18\,keV, the setup and alignment is complicated by the fact that many more Be lenses are needed, making alignment harder, and increasing the lens aberrations. In addition, the shortest focal length that can be used is generally larger, leading to a smaller magnification. Nevertheless, the higher photon energy allows using  either thicker targets, or targets containing higher Z-elements, while maintaining enough transmission to collect quality images. We tested the system with a lens stack of 98 lenses with a 50$\mu$m curvature at their apex, leading to a focal length of 266mm and an magnification of $\times$15. Measurement on resolution targets showed a resolution similar to 8.2keV. At 18keV, this is significantly larger than the theoretical diffraction limit, due to a combination of both geometric and chromatic aberration of the Be lenses.

Some examples of the use of this imaging geometry described below. In Figure~\ref{fig:shock_Si} a shockwave in a silicon sample, driven by the MEC long pulse laser is depicted. Both an elastic precursor, and phase change to can be clearly distinguished.

\begin{figure}
\includegraphics[width=\linewidth]{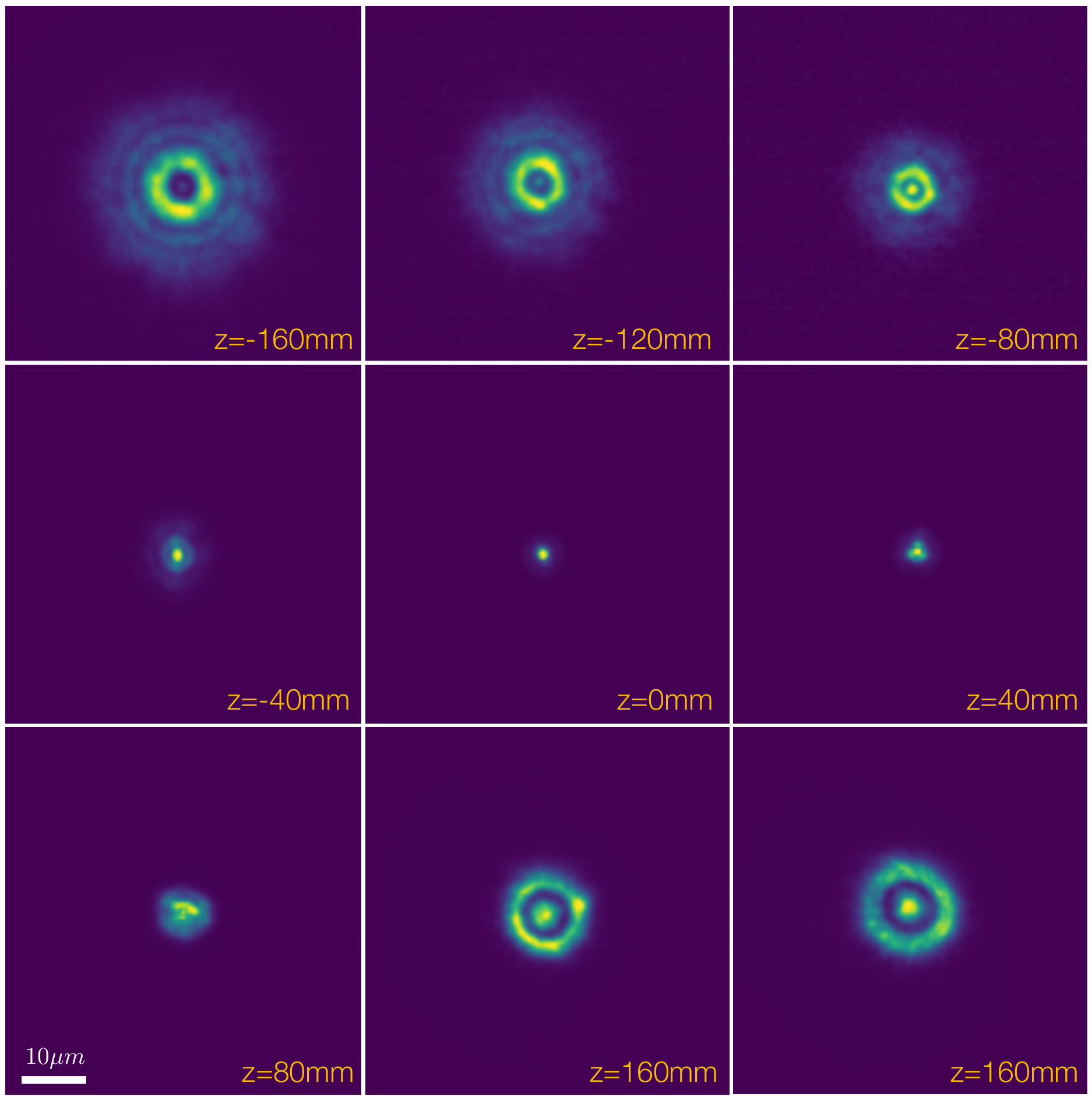}
\caption{\label{fig:FocalScan} 
Intensity profile of the focus of the MEC beamline lenses in steps of 40mm. A set of 7 be lenses with curvature of 300$\mu$m at their apex, leading to a focal length of 4.22m was chosen. 800$\mu$m pinhole were placed in before and after the lens set to restrict the x-ray to the usefull aperture of the lenses.} 
\end{figure}

The imaging setup can also be used to measure the spotsize of the standard beryllium focusing lenses that are in the MEC beamline. These lenses sit approximately 4.2\,m upstream of the center of the MEC target chamber (see \cite{Nagler2015} for details), and therefore a stack with focal length between 4.0\,m and 4.4\,m is generally used for standard MEC experiment, with a best focus around 1.2\,$\mu$m. Wavefront measurement have in the past been performed of such lens stack, showing significant spherical aberrations, generating ring-patterns around the focus~\cite{Seaberg2019}. The MXI direct imaging setup can  be used to directly image the intensity profile of the beam on the sample. In Figure~\ref{fig:FocalScan} we show the intensity profile around the focus of our beamline lenses as imaged by the MXI instrument. The airy-like ring pattern can clearly be seen.

\ctable[caption={Summary of the specifications of the MXI.}, 
	label={tab:mxi_spec}, pos=h!]{lll}{
	\tnote{At reduced magnification}
	}{
	\FL	
	    Parameter~\tmark        & Design goal               & Unit
	\ML
        Photon Energy Range     & 5 -- 24~\tmark            & keV\NN
        Repetition Rate         & On demand - 120           & Hz\NN
        Spot Size (SASE bandwidth) & 1 -- 100               & $\mu$m\NN
        Spot Size (monochromatic bandwidth) & 0.13 -- 100   & $\mu$m\NN
        Pulse Duration          & $<$ 20 -- 50              & fs
    \LL
}


\section{Talbot  Imaging }
Talbot shearing interferometry has been widely used at FELs to determine the wavefront of the of the x-ray beam, and  its focal properties in a single shot~\cite{Seaberg2019,Liu:18,Matsuyama:12}. The method uses a two dimensional phase diffraction grating placed in the beam. At a specific distance behind the grating an 'image' of the grating is reproduced. Analysis of the distortions, typically using Fourier methods~\cite{Takeda:82} yields the differential phase in two orthogonal dimensions. These differential phases can be integrated to yield the full wavefront of the beam, and backpropagating to the focus yields the focal properties. It has been an important technique to quantify the focus in experiments where x-ray  foci of order 100\,nm were required.

\begin{figure}
\includegraphics[width=\linewidth]{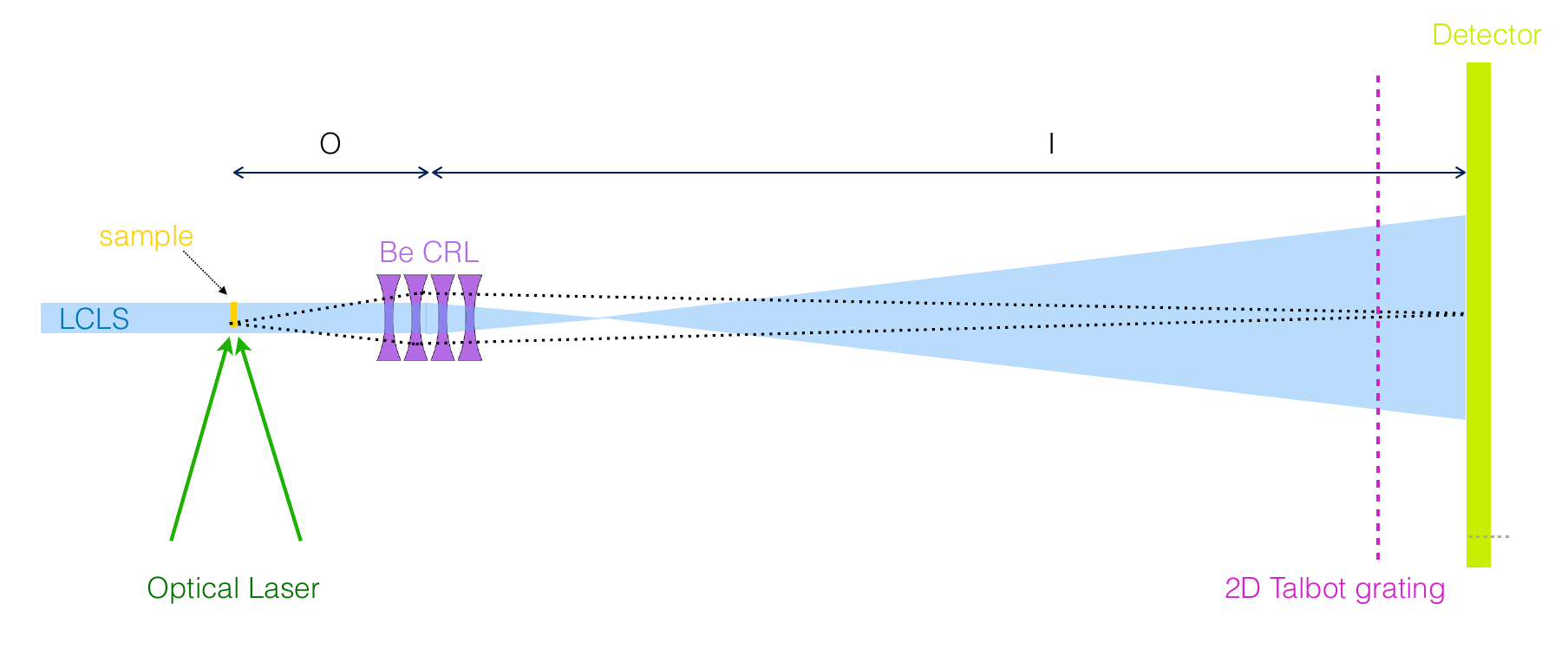}
\caption{Talbot Phase Imaging at MEC. The MXI is placed in a direct imaging geometry, where the Be lens stack images the sample onto an X-ray camera. A 2 dimensional checkerboard $\pi$-phase grating is placed at the distance of the first talbot order before the camera. \label{fig:talbotsetup} 
} 
\end{figure}

We have have combined this Talbot phase measurement technique with the direct imaging. The setup, which can be seen in Figure~\ref{fig:talbotsetup}, is essentially the same as the direct imaging setup, with the addition of a Talbot grating.  A checkerboard $\pi$-phase grating with a pitch of 22.5\,$\mu$m was placed at the first Talbot distance in front of the imaging camera. 

\begin{figure}
\includegraphics[width=\linewidth]{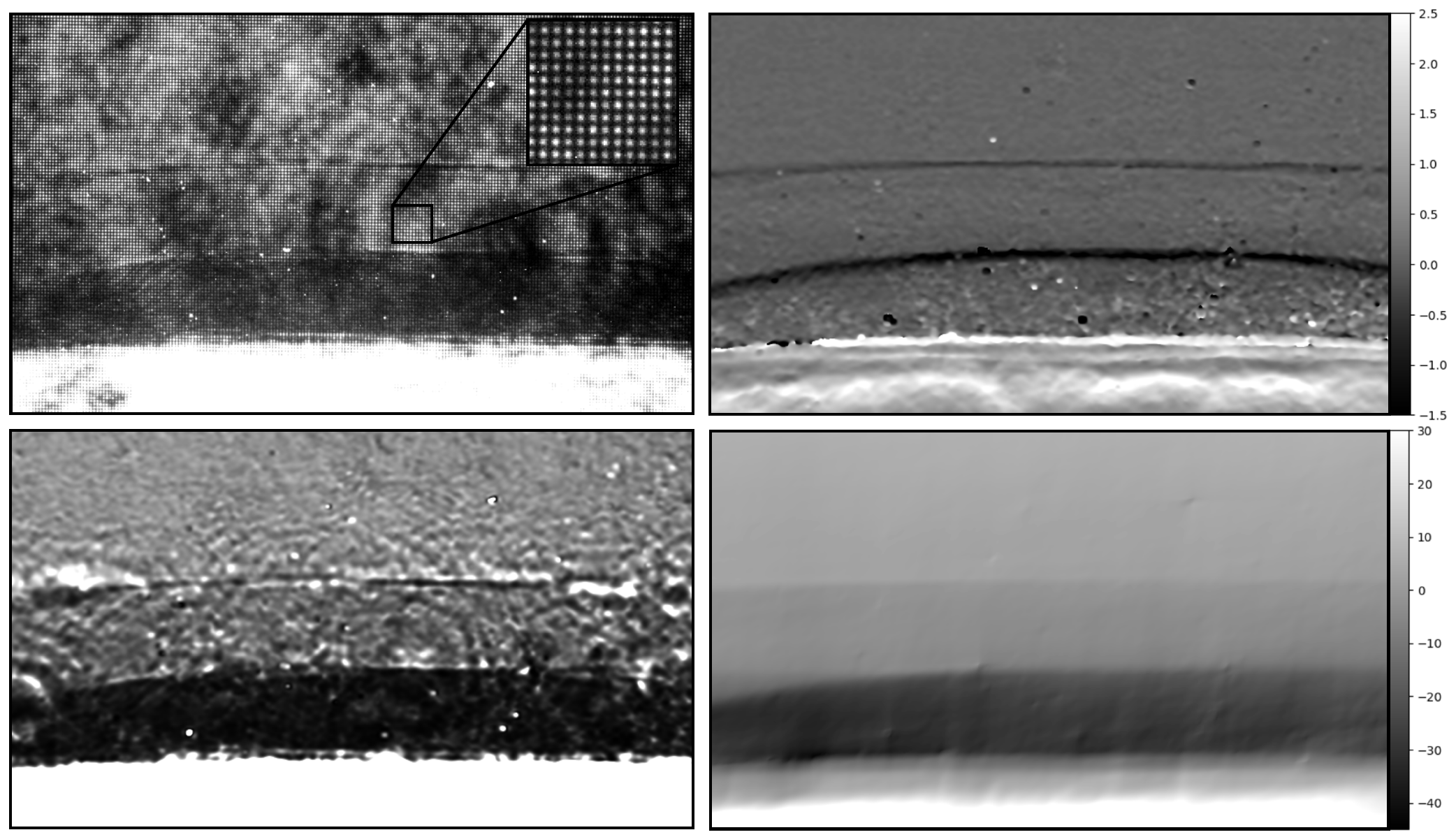}
\caption{a) Raw Talbot phase image of a shock in a silicon sample b) Differential phase in y. c) Recovered intensity change. d) Integrated phase of the sample. \label{fig:talbot} 
} 
\end{figure}

The raw image of the a shock propagating through Silicon with this setup is shown in Figure~\ref{fig:talbot}.a).  Both the elastic precursor, as well as the phase transformation to denser shocked state of Si, can be seen in the raw image. Overlayed on this image is the grid pattern introduced by the Talbot grating (an zoom region is shown in the inset) that is used to measure the phase. The 2-dimensional fringe pattern is analysed by standard Fourier methods~\cite{Takeda:82}. Since Talbot interferometry is a shearing interferometry method, the Fourier analysis results in the differential phase in the horizontal or vertical direction  depending on which Fourier peak is chosen. 

There are many ways~\cite{Takeda:82,Seaberg2019,Nagler2017,Makita:20} to integrate the differential phase and get the phase $\phi(x,y)$. Though not necessary the best, in this paper we use a fast and easy method~\cite{Morgan2012}. 
Since at the first Talbot order the shear is small, we can approximate this as:
\begin{align}
  S_x(x,y)&=\phi(x,y)-\phi(x-s,y)=s\frac{\partial \phi}{\partial x}\\
  S_y(x,y)&=\phi(x,y)-\phi(x,y-s)=s\frac{\partial \phi}{\partial y}\\
 \end{align}

 We then use  the Fourier relation:
 \begin{equation}
  \mathcal{F}( \frac{\partial \phi}{\partial x})=ik_x \mathcal{F}(
  \phi)
\end{equation}
From this follows:
\begin{equation}
  \phi=\mathcal{F}^{-1}\left(\frac{\mathcal{F} (S_x+i
      S_y)}{ik_x- k_y}    \right)
\end{equation}
which can be efficiently implement using a fast Fourier transform and its inverse.
 
We tested this method on a shock compressed silicon sample driven by the MEC long pulse laser. First an image was taken of the illumination (i.e., the beam without a sample in place) to use as a a reference.  Then a measurement of the shock in the sample was taken. Figure~\ref{fig:talbot}.a) shows the raw data of the shock, while b) shows the differential phase in the y direction. The reconstructed intensity is shown in Figure\ref{fig:talbot}.c). The structure of the Be lenses is highly visible in that image. 
In Figure\ref{fig:talbot}.d) we show the actual phase introduced by the sample; the structure  in the intensity introduced by the lenses disappears, as it is not present in the phase. The phase change behind the elastic wave can clearly be seen, as well as the larger shift after the phase change.

The Fourier analysis of the initial 2d fringe pattern limits the resolution that can be attained due to the Nyquist–Shannon sampling theorem.  In the data shown in Figure~\ref{fig:talbot} the resolution of the phase measurement is approximately 1.2$\,\mu$m. However, the imaging geometry at MEC has the capability of magnification of up to 40x at photon energies below 10\,keV. Since we can easily resolve a 4\,um Talbot pitch on the scintillator based detector, a resolution of 8\,um in the phase measurement at the detector or $\tfrac{8\,\mu m}{40}=200$~nm is  achievable.


\section{Concluding remarks}
\label{sec:conclusions}
In conclusion, we have developed at X-ray imaging diagnostic at MEC that is available to the user community. It can be used in combination with the MEC long pulse laser to, for example, image shock waves and phase transformation in materials, and with the MEC short pulse laser system, to image its interaction with dense plasmas. The diagnostic has a 200\,nm resolution, and can operate using photon energies ranging from 6keV to 24\,keV. In combination with with Talbot grating, a Talbot Imaging setup can be used to measure both intensity and phase changes introduced by a target with 200\,nm resolution.

\begin{acknowledgments}
A.E.G. acknowledges support from the LANL Reines LDRD and from the DOE FES ECA, 2019.  Use of the Linac Coherent Light Source (LCLS), SLAC National Accelerator Laboratory, is supported by the U.S. Department of Energy, Office of Science, Office of Basic Energy Sciences under Contract No. DE-AC02-76SF00515. The MEC instrument and B.N., E.G., D.K., G.D. and H.J.L are supported by the U.S. Department of Energy, Office of Science, Office of Fusion Energy Sciences under contract DE-AC02-76SF00515. 
\end{acknowledgments}


\bibliography{mxi}

\end{document}